\documentclass[12pt]{article}
\pdfoutput=1
\usepackage{putex}
\usepackage{graphicx}
\usepackage{epstopdf}
\usepackage{enumerate}
\usepackage{cite}
\usepackage{tensor}
\usepackage{slashed}

\numberwithin{equation}{section}

\newcommand{\abs}[1]{\left\lvert #1 \right\rvert}

\newcommand {\be} {\begin {equation}}
\newcommand {\ee} {\end {equation}}

\newcommand {\bes} {\begin {equation*}}
\newcommand {\ees} {\end {equation*}}

\newcommand{\es}[2] {\begin{equation} \label{#1} \begin{split} #2 \end{split} \end{equation}}

\newcommand{\CP}{\mathbb{CP}}
\newcommand{\Z}{\mathbb{Z}}

\newcommand{\R}{\mathbb{R}}

\def\Tr{\mop{Tr}}

\newcommand{\beq}{\begin{equation}}
\newcommand{\eeq}{\end{equation}}

\begin{document}

\preprint{MIT-CTP-4387}

\institution{Harvard}{Center for the Fundamental Laws of Nature, Harvard University, Cambridge, MA 02138}
\institution{MIT}{Center for Theoretical Physics, Massachusetts Institute of Technology, Cambridge, MA 02139}

\title{Exact results for five-dimensional superconformal field theories with gravity duals}

\authors{Daniel L.~Jafferis\worksat{\Harvard} and
  Silviu S.~Pufu\worksat{\MIT}}

\abstract{We apply the technique of supersymmetric localization to exactly compute the $S^5$ partition function of several large $N$ superconformal field theories in five dimensions that have $AdS_6$ duals in massive type IIA supergravity.   The localization computations are performed in the non-renormalizable effective field theories obtained through relevant deformations of the UV superconformal field theories.  We compare the $S^5$ free energy to a holographic computation of entanglement entropy in the $AdS_6$ duals and find perfect agreement.  In particular, we reproduce the $N^{5/2}$ scaling of the $S^5$ free energy that was expected from supergravity.
}
\date{July 2012}

\maketitle


\section{Introduction and overview}

Supersymmetric localization is a useful tool that makes possible exact computations in superconformal field theories (SCFTs) even at strong coupling.  The idea \cite{Witten:1988ze} is simple:  pick a supercharge $Q$ and add to the Lagrangian a $Q$-exact operator whose bosonic part is positive-definite.  This change in the action does not affect the value of the partition function or of correlation functions of any number of $Q$-invariant operators.  One can then evaluate these quantities exactly by taking the coefficient of the $Q$-exact term to be large and using a saddle point approximation.

In favorable situations, the solution space of the saddle point equations is finite dimensional, typically involving configurations on the Coulomb branch in which the fields in the Lagrangian are constant throughout space. This procedure reduces the infinite-dimensional path integral to a manageable finite-dimensional integral.  For each saddle, the integrand in the finite-dimensional integral comes from the one-loop determinant of fluctuations around the saddle.  By taking the coefficient of the $Q$-exact term to infinity, it can be argued that even though the finite-dimensional integral was derived from a saddle point approximation, it actually provides an exact result.

Localization was used to find the exact partition function of 3d SCFTs on the three-sphere in \cite{Kapustin:2009kz, Jafferis:2010un, Hama:2010av} for theories with ${\cal N} \geq 2$ supersymmetry, building up on the 4d results of \cite{Pestun:2007rz}.  One of the remarkable results that followed was the evaluation of the free energy on $S^3$, defined as $F = - \log \abs{Z_{S^3}}$, for SCFTs with gravity duals, the goal being to test the AdS/CFT duality \cite{Maldacena:1997re, Gubser:1998bc, Witten:1998qj}.  Many such SCFTs can be realized as effective theories on $N$ coincident M2-branes placed at the tips of Calabi-Yau cones \cite{Aharony:2008ug, Jafferis:2008qz, Jafferis:2009th, Benini:2009qs, Martelli:2009ga}.  From the gravity side of the AdS/CFT duality one expects $F \propto N^{3/2}$ at large $N$, the fractional power of $N$ having been regarded as a puzzle.\footnote{The $N^{3/2}$ scaling was first noticed in \cite{Klebanov:1996un} for the thermal free energy of $N$ M2-branes in flat space.}  Using the localization results of \cite{Kapustin:2009kz, Jafferis:2010un, Hama:2010av} in a large class of such theories, it was possible to match the $S^3$ free energy predicted from the gravity side with a field theory computation, and in particular to reproduce the $N^{3/2}$ scaling using just field theory methods \cite{Drukker:2010nc, Herzog:2010hf, Jafferis:2011zi, Martelli:2011qj, Cheon:2011vi}.

In this paper we aim to obtain similar results for
five-dimensional SCFTs with gravity duals.\footnote{See also \cite{Kallen:2012zn}, where supersymmetric (but not conformal) theories on $S^5$ were studied in relation to the $(2, 0)$ theory in six dimensions.}  Localization on $S^5$
for theories with ${\cal N} = 1$ SUSY was partially worked out in
\cite{Kallen:2012cs, Hosomichi:2012ek, Kallen:2012va}, and ${\cal
N} = 1$ SCFTs with supergravity duals were originally proposed in
\cite{Intriligator:1997pq, Brandhuber:1999np} (see also
\cite{Seiberg:1996bd, Morrison:1996xf}) and generalized recently
in \cite{Bergman:2012kr} to quiver-type theories.  The stringy
origin of these field theories is in type I' string theory as
strongly-coupled microscopic theories on the intersection of $N$
D4-branes and some number of D8-branes and orientifold planes.
From the massive type IIA supergravity backgrounds dual to the UV
SCFTs one expects the number of degrees of freedom to scale as
$N^{5/2}$ at large $N$ \cite{Brandhuber:1999np}, but explicit
formulas for the free energy on $S^5$, $F = - \log \abs{Z_{S_5}}$,
have not been determined.

Unlike in 3d where the strongly-coupled SCFT is in the deep IR,
and one can perform localization in a weakly-coupled UV theory
that flows in the IR to the SCFT of interest, in 5d the strongly-coupled SCFT is at the UV fixed point.  Each such UV theory has a
relevant deformation that makes it flow in the IR to free
Yang-Mills theory coupled to matter, which is a non-renormalizable field theory
because in five dimensions the Yang-Mills coupling $g_\text{YM}^2$ has dimensions of $1/$mass.

One can think of defining the Yang-Mills-matter theory at some
scale $\Lambda$ where one should also specify an infinite set of
irrelevant operators that are needed to describe the UV fixed
point with arbitrary accuracy at sufficiently large $\Lambda$. We expect that the gauge invariant supersymmetric
irrelevant operators beyond the Yang-Mills term are $Q$-exact, so we can tune their coefficients at
will without changing the path integral on $S^5$.  The standard
rotationally symmetric Yang-Mills Lagrangian itself fails to be
$Q$-exact \cite{Hosomichi:2012ek}.
The partition
function on $S^5$ thus depends on the dimensionless parameter
$r/g_\text{YM}^2$, where $r$ is the radius of the $S^5$,
describing the relevant deformation of the UV SCFT.

Independently of this non-trivial Yang-Mills term, when performing
localization, a $Q$-exact kinetic term for the gauge fields is
also added. This term, which breaks rotational invariance, does
not change the value of the path integral, but it renders the
theory free in the IR\@. The localization computation is therefore
performed in the Yang-Mills-matter theory, far from the UV fixed
point.   We can extract the UV SCFT free energy on $S^5$ from the
Yang-Mills-matter theory provided that no new operators beyond the
$Q$-exact irrelevant operators  are important at large energy
scales.  This is precisely the situation found in
\cite{Seiberg:1996bd}, since no new heavy
fields appear at higher energies---one simply tunes the irrelevant
operators that correct the 5d Yang-Mills matter theory.  From the point of view of the UV SCFTs, which in this
case have $E_n$ global symmetry,\footnote{The $E_n$ global
symmetry enhancement was confirmed in \cite{Kim:2012gu} by
calculating the superconformal index of these theories on $S^4
\times S^1$ using localization.} there exist $n$ independent
deformation parameters $m_i$ analogous to the real mass parameters
of 3d field theories.  They are the constant values of the real scalar
in a background vector multiplet that couples to the global $E_n$
current.  One of these parameters generates a flow to the
Yang-Mills-matter theory, while the other ones generate flows
between the $E_n$ SCFTs by decreasing $n$.

There are two potential difficulties in extending the successful
AdS/CFT tests from three to five dimensions, one regarding the
field theory localization computation and one regarding the dual
supergravity description.  The first is that on $S^5$ the gauge
field localizes on classical configurations involving instantons
whose understanding is currently incomplete.  It is only the
contribution from the sector with no instantons that has so far
been calculated \cite{Hosomichi:2012ek, Kallen:2012cs,
Kallen:2012va}. However, as we show in the next section, this
difficulty is surmounted by taking the large $N$ limit where one
can argue that the contributions from the sectors with instantons
should be suppressed.

The second difficulty present in our setup is that  the dual
supergravity background has curvature singularities.  While at
large $N$ the supergravity approximation is reliable arbitrarily
close to the singularity, a naive evaluation of the on-shell
action results in a divergent $S^5$ free energy.  We overcome this
difficulty by using the result of \cite{Casini:2011kv} where it
was shown that in a CFT the free energy on $S^5$ equals the
universal part of the entanglement entropy across a 3-sphere in
flat Minkowski spacetime.  In turn, the entanglement entropy can
be computed holographically using the generalization of the
proposal of \cite{Ryu:2006bv} to setups with varying dilaton
\cite{Klebanov:2007ws}.  It turns out that this entanglement
entropy computation is not plagued by the divergences that arise
when trying to evaluate the on-shell action.  Moreover, it is an
inherently simpler computation because it involves just the metric
and the dilaton as opposed to all supergravity fields.

We find agreement between the $S^5$ free energy computed in the field theory using localization and the entanglement entropy computed on the gravity side.  Our main result is that the $S^5$ free energy, computed either from field theory or gravity, is
 \es{FPreview}{
  F = - \frac{9 \sqrt{2} \pi n^{3/2} N^{5/2}}{5 \sqrt{8 - N_f}} + o(N^{5/2}) \,,
 }
for the class of $\Z_n$ orbifold theories engineered in type I'
string theory from $N$ D4-branes, $N_f$ D8-branes, and one
O8-plane that were described in \cite{Bergman:2012kr}.  This
result is a check on several other results/conjectures:  the
AdS/CFT relation between the field theory \cite{Intriligator:1997pq, Aharony:1997ju, Aharony:1997bh} and gravity backgrounds
\cite{Brandhuber:1999np, Bergman:2012kr}, the equality between the
free energy on $S^5$ and the entanglement entropy across $S^3$
\cite{Casini:2011kv}, and the holographic prescription for
computing the entanglement entropy \cite{Ryu:2006bv,
Klebanov:2007ws}.

One could conjecture that the quantity $-F$ is positive for all 5d conformal field theories, and that $-F_{\text{UV}} > - F_{\text{IR}}$ for any renormalization group flow, in analogy with the $F$-theorem proposed in three dimensions \cite{Myers:2010tj, Jafferis:2011zi, Klebanov:2011gs, Casini:2012ei}.  This statement is consistent with the observation that in theories with holographic duals, the entanglement entropy across a three-sphere decreases under RG flow \cite{Myers:2010tj,Myers:2010xs}, and at the endpoints of the flow it approaches $-F$ \cite{Casini:2011kv}.  It is also consistent with the observation that $-F$ decreases under slightly relevant perturbations of a CFT, and that it remains constant under exactly marginal deformations \cite{Klebanov:2011gs}.  In the examples of \cite{Brandhuber:1999np, Bergman:2012kr}, for which $F$ is given by \eqref{FPreview}, there are RG flows under which the number of flavors $N_f$ decreases.\footnote{From the point of view of the type I' brane construction these flows correspond to moving away some number of D8-branes such that they no longer intersect the D4-branes.  Similar RG flows in 3d theories realized on brane intersections were studied in \cite{Gulotta:2011si}.}  Our result \eqref{FPreview} is consistent with $-F_{\text{UV}} > - F_{\text{IR}}$ in these RG flows.

The rest of this paper is organized as follows.  In the next section we review the 5d localization results and explain why we think the instantons should be suppressed in the large $N$ limit.  In section~\ref{MATRIX} we calculate the $S^5$ partition function at large $N$ for all the theories whose gravity duals were described in \cite{Brandhuber:1999np, Bergman:2012kr}.  In section~\ref{ENTANGLEMENT} we present the entanglement entropy computation for the dual gravity solutions.

\section{Exact 5d SCFT $S^5$ partition function from the IR
Lagrangian}

\subsection{Supersymmetric localization results}

The five-dimensional superconformal theories we will examine lack
a Lagrangian description at the conformal fixed point. However,
one may deform these theories by a relevant operator, such that in
the IR they flow to 5d ${\cal N}=1$ Yang-Mills coupled to various
hyper multiplets \cite{Seiberg:1996bd}.   As shown in
\cite{Kallen:2012va} (see \cite{Hosomichi:2012ek, Kallen:2012cs}
for earlier work and also \cite{Kim:2012av}), the $S^5$ partition
function of 5d Yang-Mills theory with gauge group $G$ and matter
hyper multiplets in representations $R_i$ takes the form
 \es{PartFunction}{
  Z &= \frac{1}{\abs{\cal W} } \int_\text{Cartan}d \sigma \, e^{- \frac{4 \pi^3 r}{g_\text{YM}^2} \tr_F \sigma^2
   + \frac{\pi k}{3} \tr_F \sigma^3 }
   \text{det}_\text{Ad}  \left(\sin(i \pi \sigma) e^{\frac 12 f(i \sigma)}  \right) \\
    &\times \prod_I \text{det}_{R_I} \left(\cos (i \pi \sigma) e^{-\frac 14 f\left(\frac 12 - i \sigma\right) - \frac 14 f\left(\frac 12 + i \sigma\right)} \right) + \text{instanton contributions} \,,
 }
where $g_\text{YM}^2$ is the bare Yang-Mills coupling, $k$ is a possible Chern-Simons level, ${\cal W}$ is the Weyl group of $G$, and $\tr_R$ and $\det_R$ are, respectively, the trace and the determinant in representation $R$, $F$ being the fundamental representation and Ad being the adjoint.  The function $f$ is defined as
 \es{fDef}{
  f(y) \equiv \frac{i \pi y^3}{3} + y^2 \log \left( 1- e^{- 2 \pi i y} \right) + \frac{i y}{\pi} \text{Li}_2 \left(e^{- 2 \pi i y} \right)
   + \frac{1}{2 \pi^2} \text{Li}_3 \left( e^{-2 \pi i y} \right) - \frac{\zeta(3)}{2 \pi^2} \,.
 }
This Yang-Mills matter theory is free at low energies, and the flow is associated to the 5d Yang-Mills term, $\frac{1}{2 g_\text{YM}^2} \int \Tr(F^{\mu \nu} F_{\mu \nu})$, which is an irrelevant operator from the IR point of view.  The UV CFT has a moduli space of vacua, which maps to the Coulomb branch of the IR Yang-Mills theory.  The integration variables $\sigma$ appearing in \eqref{PartFunction} are precisely the parameters describing the Coulomb branch:  they are the expectation values, in the Cartan of the gauge group, of the real scalar in the vector multiplet.

The partition function \eqref{PartFunction} contains some information that is not new to putting the Yang-Mills-matter theory on $S^5$.  To see this, it is convenient to write
 \es{ZSaddle}{
  Z &= \frac{1}{\abs{\cal W}} \int_\text{Cartan} d\sigma\, e^{- F(\sigma)} \,, \\
   F(\sigma) &\equiv \frac{4 \pi^3 r}{g_\text{YM}^2} \tr_F \sigma^2
      + \frac{\pi k}{3} \tr_F \sigma^3
      + \tr_\text{Ad} F_V(\sigma) + \sum_I \tr_{R_I}F_H(\sigma) \,,
 }
where the functions $F_V(\sigma)$ and $F_H(\sigma)$ can be easily read off from \eqref{PartFunction}.  By expanding these functions at large arguments we see that
 \es{FVH}{
  F_V(y) &\approx \frac{\pi}{6} \abs{y}^3 - \pi \abs{y}  \,, \\
  F_H(y) &\approx -\frac{\pi}{6} \abs{y}^3 - \frac{\pi}{8} \abs{y}\,,
 }
provided $\abs{y} \gg 1$, which is the only information about these functions that will be needed for the rest of this paper.  We then obtain
 \es{FLargesigma}{
  F(\sigma) =  \frac{4 \pi^3 r}{g_\text{YM}^2} \tr_F \sigma^2 + \frac{\pi}{3}
    \left[k \tr_F \sigma^3 +  \frac 12 \tr_\text{Ad} \abs{\sigma}^3 - \frac 12 \sum_I \tr_{R_I} \abs{\sigma}^3 \right] + O(\abs{\sigma}) \,.
 }
If ${\cal F}$ is the prepotential of the Yang-Mills-matter theory, which we write as
 \es{Prepot}{
  {\cal F} (\sigma)= \frac 12 h_{ij} \lambda_i \lambda_j  + \frac{c_{ijk}}{6} \lambda_i \lambda_j \lambda_k \,,
 }
$\lambda_i$ being the components of $\sigma$ in the Cartan basis, one can check that
 \es{FPrepot}{
  F(\sigma ) \propto h_{ij} \lambda_i \lambda_j  + c_{ijk} \lambda_i \lambda_j \lambda_k \,.
 }
 The function $F(\sigma)$ therefore knows about the $U(1)^r$ gauge theory on the Coulomb branch, where $r$ is the rank of the gauge group.  In particular, it encodes the effective CS level and the effective gauge coupling, which are given, respectively, by the third and second derivatives of the prepotential.    As a check, note that integrating out an odd number of fermions far on the Coulomb branch produces a half-integral CS level, and indeed, the last two terms in \eqref{FLargesigma} come with a relative factor of $1/2$ compared to the Chern-Simons term.  From now on we set the CS level $k=0$.

\subsection{Simplifications in the large $N$ limit}

We now argue that in computing \eqref{PartFunction} in the supergravity regime for the field theories with gravity duals of \cite{Bergman:2012kr} one can ignore both the first factor in the integral \eqref{PartFunction} as well as the instanton contributions.  Let us illustrate our reasoning in the case of the simplest theory we will consider, namely the $USp(2N)$ Yang-Mills theory originally introduced in \cite{Intriligator:1997pq} with matter consisting of a hyper multiplet in the antisymmetric representation of the gauge group and $N_f$ fundamental hyper multiplets.

The Coulomb branch is parameterized by the scalar in the vector
multiplet having a nonzero expectation value in the Cartan of the
gauge group, $\sigma = \diag\{\lambda_1, \ldots, \lambda_N,
-\lambda_1, \ldots, -\lambda_N \}$.   The moduli space is
quotiented by the action of the Weyl group which sends $\lambda_i
\to -\lambda_i$ for each $i$ independently and permutes the
$\lambda_i$.  Instead of restricting the $\lambda_i$ to a Weyl
chamber, in writing \eqref{PartFunction} we chose to let the
$\lambda_i$ be unrestricted and divided the partition function by
a factor of $\abs{\cal W} = 2^N N!$.   On the Coulomb
branch the gauge group is broken down to $U(1)^N$, and there is a
one-loop correction to the effective gauge coupling for the $i$th
$U(1)$ factor \cite{Witten:1996qb} that can be found from \eqref{FLargesigma}:
 \es{effectiveGauge}{
  \frac{r}{g_{\text{eff}, i}^2(\sigma)} = \frac{r}{g_\text{YM}^2} + \frac{1}{12 \pi^2} (8-N_f) \abs{\lambda_i} \,.
 }
In \cite{Seiberg:1996bd, Intriligator:1997pq}, it was argued that when $N_f < 8$, the effective coupling on the moduli space remains finite even when $g_\text{YM}^2 \rightarrow \infty$, and there exists a UV CFT at the origin where $\lambda_i = 0$ for all $i$.  To access the fixed point we should just set $r/g_\text{YM}^2 = 0$.

We claim that the Yang-Mills contribution $r \tr \sigma^2 /
g_\text{YM}^2$ can be
ignored provided $r/g_\text{YM}^2 \ll \sqrt{N}$. Indeed, in the
next section we will find that the main contribution to the
zero-instanton integral in \eqref{PartFunction} comes from
configurations where $\sigma = O(\sqrt{N})$.  For these
configurations, the logarithm of the product of determinants is of
order $N^{5/2}$ while the Yang-Mills term is of order $r N^2 /
{g_\text{YM}^2}$.  Hence the the Yang-Mills contribution can be
ignored if $r/g_\text{YM}^2 \ll \sqrt{N}$.

In the dual geometry, adding the operator that makes the UV SCFT flow to Yang-Mills matter theory in the IR corresponds to turning
on a light scalar field. One would have expected that when $g_\text{YM}^2$ was of order one in units of $r$, the radius of the
$S^5$, the geometric description would break down in the interior at some distance of order one in AdS units, since this relevant deformation results in weakly coupled field theory at low energies. It would be interesting to understand why this does {\it not} appear to result in any change in the sphere free energy at leading order.

Arguing that the instanton contributions can also be ignored at large $N$ is slightly more subtle.  First, let's show that they can definitely be ignored when $r/g_\text{YM}^2 \gg 1$.  Given that we eventually want to be able to send $r/g_\text{YM}^2$ to zero, we will later have to refine our argument.

The instanton contribution that we suppressed when we wrote \eqref{PartFunction} consists of a sum over instantons satisfying the equations $v_\mu F_{\nu \rho} \epsilon^{\mu \nu \rho \sigma \tau} = F^{\sigma \tau}$ and $v^\mu F_{\mu \nu}=0$, where $v_\mu$ is a vector field that generates a freely acting $U(1)$ isometry of $S^5$ for which the quotient space is $\CP^2$. These equations roughly describe a self-dual instanton on $\CP^2$ smeared over the $S^1$ fiber.  For each instanton, the real scalar in the vector multiplet is covariantly constant at
the $Q$-fixed loci, while all of the fields in hyper multiplets must vanish.  Just as in the zero-instanton sector that was written explicitly in \eqref{PartFunction}, for each such configuration one has to calculate a one-loop determinant of fluctuations, and after the addition of the $Q$-exact localizing terms the one-loop approximation becomes exact.  The bottom line is that each term in the instanton sum is similar, but not necessarily identical, to the zero-instanton term written down explicitly in \eqref{PartFunction}, and it is multiplied by $e^{-S_\text{inst}}$, where $S_\text{inst}$ is the classical action of the instanton configuration.

For the theories we will discuss, only the Yang-Mills term contributes to $S_\text{inst}$.   This contribution fails to be $Q$-exact by a term given
by \cite{Hosomichi:2012ek}
 \es{NotExact}{
 S_\text{inst} = \frac{1}{g_\text{YM}^2} \int v \wedge \Tr( F \wedge F) + \text{SUSY completion} \,,
}
which counts the number of instantons.  Note that the supersymmetrization of the first term in \eqref{NotExact} includes the quadratic term for the real scalar that appears in the localized matrix integral and gives rise to the quadratic term in \eqref{PartFunction}.  Each instanton configuration is therefore weighted by $e^{-I r / g_\text{YM}^2}$, where $I$ is the instanton number.

The sum over instantons would seem to play an important role in describing the UV physics, since they are exponentially suppressed only by the classical action $\exp (- S_\text{inst} )$.  This exponential factor goes to $1$ at the UV conformal point where $r/g_\text{YM}^2 = 0$, and there the instantons would not be suppressed at all.  The instantons are exponentially suppressed if $r/ g_\text{YM}^2 \gg 1$.  In particular, they would be exponentially suppressed if $g_\text{YM}^2 \rightarrow 0$, namely the limit of weakly coupled 5d Yang-Mills theory, far from the UV fixed point.

As mentioned above, in the next section we find that the zero-instanton matrix integral is surprisingly independent of the Yang-Mills term as long as $r/g_\text{YM}^2 \ll N^{1/2}$, so there is a
regime $1 \ll r/g_\text{YM}^2 \ll \sqrt{N}$ where the analysis in the next section, which neglects both the instanton sum and the perturbative
Yang-Mills term, is clearly valid. In terms of the 't Hooft coupling $t \equiv N g_\text{YM}^2 / r$, this regime is $\sqrt{N} \ll t \ll N$.

To directly compare to $AdS_6$, however, we need to turn off the
relevant deformation completely by sending $r / g_\text{YM}^2 \rightarrow 0$, so we do not want to be stuck working in the regime $1 \ll r/g_\text{YM}^2 \ll \sqrt{N}$. As explained in \cite{Seiberg:1996bd}, the parameter that controls
the instanton-soliton expansion is the central charge given schematically by
 \es{CentralCharge}{
  Z  \sim I_3 \abs{\lambda} + \left( 2 (8 - N_f) \abs{\lambda} + \frac{r}{g_\text{YM}^2}  \right) I \,,
  }
where $I_3$ is the electric charge and $I$ is as before the instanton-soliton number.  In other words, it is really the effective Yang-Mills coupling $r/g_\text{eff}^2$ in \eqref{effectiveGauge} that weights the instanton action as opposed to just $r/g_\text{YM}^2$ as one would naively infer from \eqref{NotExact}.  This weight depends on the Coulomb branch parameters $\lambda_i$.

To see this from the localized path integral, one should compute
the one-loop determinants of the vector and hyper multiplets in the
instanton background, and observe a correction proportional to the
instanton number. It would be interesting to do this explicitly.
Note that the one-loop shift $g_\text{YM}^2 \to g_\text{eff}^2$ is easily seen in the leading term \eqref{FLargesigma}, and we claim that such a shift should also occur for the $I r / g_\text{YM}^2$ term that appears in non-trivial instanton backgrounds.

As shown in the next section, the large $N$ saddle point equations imply that the eigenvalues are spread in a clump of size $O(\sqrt{N})$ on the Coulomb branch. Therefore, the contributions of instantons are indeed exponentially suppressed in $r/{g_\text{eff}^2} \sim \sqrt{N}$ even when $r/{g_\text{YM}^2} = 0$.  This suppression justifies ignoring the instanton contributions.  These contributions are important only at the origin of moduli space, but this is essentially a measure zero subset of the integration range in \eqref{PartFunction}.

In the quiver gauge theories, the one-loop contribution from bifundamental hyper multiplets drives $1/g_\text{eff}^2$ for one of the gauge groups to negative values as one moves out on the Coulomb branch associated to the other gauge group.  The effective gauge couplings only
remain finite in some region of the moduli space, hence
it was believed that such theories could not lead to UV SCFTs \cite{Intriligator:1997pq}.
However, based on a IIB brane construction involving $(p, q)$ fivebranes, it was argued in\cite{Aharony:1997ju, Aharony:1997bh} that as the
effective coupling grows large in these models, one can switch to
an s-dual description. The  $(p,q)$ fivebranes wrap a common 4+1 dimensions, where the low energy field theory lives, and a one-dimensional web in $\mathbb{R}^2$. The s-duality of IIB together with a flip in the plane containing the web switches the role of the inverse gauge coupling and the Coulomb branch parameters in the 5d theory.

As an illustrative example, consider the $SU(2) \times SU(2)$ gauge theory with a bifundamental hyper multiplet. The Coulomb branch is parameterized by the real scalars $\lambda$ and $\tilde\lambda$ in the Cartan of the gauge group. On the moduli space, the gauge symmetry is broken down to $U(1) \times U(1)$.  From the prepotential, one can derive that the matrix of effective gauge couplings is given by
 \es{geffMatrix}{
\frac{1}{g_\text{eff}^2} \sim \left( \begin{array}{cc} 4 |\lambda| -  |\lambda - \tilde\lambda| -|\lambda +\tilde\lambda| & -|\lambda+\tilde\lambda| +  |\lambda-\tilde\lambda| \\ -|\lambda+\tilde\lambda| +|\lambda-\tilde\lambda| & 4 |\tilde\lambda| - |\lambda-\tilde\lambda| -|\lambda+\tilde\lambda| \end{array} \right)\,.
 }
Taking $\tilde\lambda = 0$ results in a negative effective coupling for the second $U(1)$ factor. As one of the gauge couplings diverges, the instantons become important, and the instanton expansion must be resummed in the regions of negative effective coupling---the original gauge theory has become a bad description, and one should switch to the s-dual variables.

In the saddle point configurations that we find in the models examined in the next section, the eigenvalues in the different gauge groups are set equal, so this problematic region of the moduli space is avoided. Therefore, we are still justified in ignoring the contributions of instantons.

\section{Matrix model computations}
\label{MATRIX}

From now on we will ignore the perturbative Yang-Mills factor and the instanton contributions in~\eqref{PartFunction} and set $k=0$, focusing only on the two determinant factors.

We wish to evaluate \eqref{ZSaddle} in the saddle point approximation in a number of examples where the number of integration variables is large by finding configurations $\sigma = \sigma_*$ that extremize $F(\sigma)$.   As we will see, in our examples the Weyl group acts non-trivially on the set of saddle points, so the integral \eqref{ZSaddle} has $\abs{\cal W}$ distinct saddle points that give equal contributions to $Z$.  The integral is then approximated as $Z \approx e^{- F(\sigma_*)}$, where $\sigma_*$ is any one of these saddle points.

\subsection{$USp(2N)$ theory with matter}

The first example we examine is that of a $USp(2N)$ gauge theory with $N_f$ matter hyper multiplets in the fundamental representation and one hyper multiplet in the antisymmetric representation of $USp(2N)$.  There are $N$ elements in the Cartan of $USp(2N)$, which we will denote by $\lambda_i$, $1 \leq i \leq N$, and these will be our integration variables.  If we normalize the weights of the fundamental representation of $USp(2N)$ to be $\pm e_i$, the $e_i$ forming a basis of unit vectors for $\R^N$, the antisymmetric representation then has weights $e_i \pm e_j$ with $i \neq j$, and the adjoint representation has weights $e_i \pm e_j$ with $i \neq j$ as well as $\pm 2 e_i$.  Explicitly, the function $F(\lambda_i)$ becomes
 \es{traces}{
  F(\lambda_i) &= \sum_{i \neq j} \left[  F_V(\lambda_i - \lambda_j) + F_V(\lambda_i + \lambda_j)
   +  F_H(\lambda_i - \lambda_j) + F_H(\lambda_i + \lambda_j) \right]  \\
    &+ \sum_i \left[ F_V(2 \lambda_i) + F_V (-2 \lambda_i) + N_f F_H(\lambda_i) + N_f F_H(-\lambda_i) \right] \,.
 }
If we think of the $\lambda_i$ as the positions of $N$ particles, the first line should be interpreted as an interaction energy between these particles, while the second line is the energy in an external potential.

Note that there is a cancelation between the cubic interaction forces in the first line of \eqref{traces} that has its origins in the fact that $F_V(y) = -F_H(y)$ to leading order in $y$, as can be seen from \eqref{FVH}.    Also, if some configuration $\lambda_i = \lambda_{i*}$ extremizes \eqref{traces}, then we can construct many other configurations that extremize \eqref{traces} with the same $F(\lambda_{i})$ corresponding to the action of the Weyl group of $USp(2N)$.  In particular, we can permute the $\lambda_i$ or we can independently flip the sign of any given $\lambda_i$.  From now on we will restrict ourselves to finding extrema of \eqref{traces} where $\lambda_i \geq 0$ for all $i$.

Let's assume self-consistently that as we take $N \to \infty$, we have $\lambda_i = N^\alpha x_i$ with $\alpha>0$ and $x_i$ of order $O(N^0)$.  We furthermore introduce the density
 \es{DensityDef}{
  \rho(x) = \frac{1}{N} \sum_{i=1}^N \delta( x- x_i) \,,
 }
which approaches an $L^1$ function in the continuum limit $N \to \infty$ normalized so that
 \es{rhoNorm}{
  \int dx\, \rho(x) = 1\,.
 }
Since by assumption the $\lambda_i$ become large at large $N$, it is justified to use the approximations in \eqref{FVH}.  In the continuum limit \eqref{traces} becomes
 \es{FDensity}{
  F \approx - \frac{9 \pi}{8} N^{2 + \alpha} \int dx\, dy\, \rho(x) \rho(y) \left(\abs{x - y} + \abs{x+y} \right)
   + \frac{\pi (8 - N_f)}{3} N^{1 + 3 \alpha} \int dx\, \rho(x) \abs{x}^3 \,.
 }
In this expression we kept only the leading large $N$ behavior for each line in \eqref{traces}.  We can find a non-trivial saddle point provided that the powers of $N$ that appear in the two terms match, so $\alpha = 1/2$, and therefore $F \propto N^{5/2}$.

It is straightforward to show that a normalized density $\rho(x)$ that extremizes \eqref{FDensity} is
 \es{Gotrho}{
  \rho(x) = \frac{2 \abs{x}}{x_*^2} \,, \qquad
    x_*^2 = \frac{9}{2 (8 - N_f)} \,,
 }
for $x \in [0, x_*]$, and $\rho(x) = 0$ for $x$ outside this interval.   Plugging this configuration back into \eqref{FDensity} one obtains that the free energy on $S^5$ is
 \es{GotF}{
  F \approx - \frac{9 \sqrt{2} \pi N^{5/2}}{5 \sqrt{8 - N_f}} \,.
 }

\subsection{Orbifold theories}

We now extend our analysis to the $\Z_n$ orbifold theories found in \cite{Bergman:2012kr}.  There are three distinct classes of such theories, two for the case where $n$ is even and one for odd $n$.  As we will explain below, in each of these three cases the free energy functional is
 \es{FreeZn}{
  \frac{F}{N^{5/2}} \approx - \frac{9 \pi n}{8} \int dx\, dy\, \rho(x) \rho(y) \left(\abs{x - y} + \abs{x+y} \right)
   + \frac{\pi (8 - N_f)}{3} \int dx\, \rho(x) \abs{x}^3
 }
in the continuum limit.  In other words, the first term in \eqref{FDensity} got multiplied by $n$ but the second term stayed unchanged.  The on-shell free energy can be easily computed by recalculating the density $\rho(x)$ that extremizes \eqref{FreeZn}, or equivalently by using the following scaling argument.  If we send $x \to \sqrt{n} x$ and $\rho(x) \to \rho(x) / \sqrt{n}$ (so that $\rho(x)$ would still be normalized as in \eqref{rhoNorm}), \eqref{FreeZn} becomes
  \es{FreeZnAgain}{
  \frac{F}{n^{3/2} N^{5/2}} \approx - \frac{9 \pi}{8} \int dx\, dy\, \rho(x) \rho(y) \left(\abs{x - y} + \abs{x+y} \right)
   + \frac{\pi (8 - N_f)}{3} \int dx\, \rho(x) \abs{x}^3 \,.
 }
It follows right away that the free energy we found in the previous section gets multiplied by $n^{3/2}$:
 \es{FOrbi}{
  F \approx - \frac{9 \sqrt{2} \pi n^{3/2} N^{5/2}}{5 \sqrt{8 - N_f}} \,.
 }
This is our main result on orbifold theories.  In the rest of this section we explain why \eqref{FreeZn} holds in each of the three cases described in \cite{Bergman:2012kr}.

\begin{figure}[htb]
\begin{center}
\leavevmode
\newcommand{\svgwidth}{.5\textwidth}
\definecolor{orange}{RGB}{255,165,0}
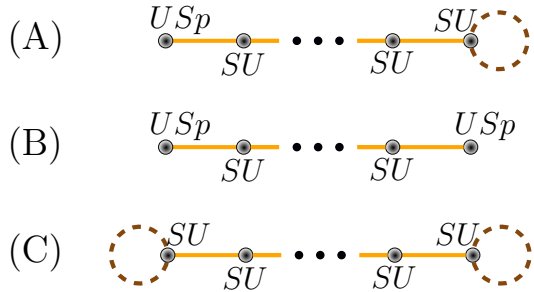
\end{center}
\caption{The three families of SCFTs discussed in \cite{Bergman:2012kr}.  The dots represent 5d $SU(2N)$ or $USp(2N)$ vector multiplets;  the solid orange lines are hyper multiplets in bifundamental representations; and the dashed brown lines are hyper multiplets in the antisymmetric representation.  In addition, one may have fundamental hyper multiplets charged under any of the gauge group factors.}
\label{Quiver}
\end{figure}

As described in \cite{Bergman:2012kr}, for each $\Z_n$ orbifold theory the gauge group is a product of $USp(2N)$ and $SU(2N)$ factors:  if $n = 2k+1$, we have $G = USp(2N) \times SU(2N)^k$, and if $n = 2k$ we either have $G = USp(2N) \times SU(2N)^{k-1} \times USp(2N)$ or $G = SU(2N)^k$.  As we will see, for the purposes of our computation, each $USp(2N)$ factor is roughly half of an $SU(2N)$ factor, so each of these theories would have roughly  $n$ $USp(2N)$ factors.  The matter consists of bifundamental hyper multiplets between adjacent gauge group factors, hyper multiplets transforming in the antisymmetric tensor representation of the outer $SU(2N)$ groups, and $N_f^{(a)}$ hyper multiplets in the fundamental representation of the $a$th gauge group factor.  See figure~\ref{Quiver}.

The analog of \eqref{traces} can be written explicitly in each of the three cases described above.  It is convenient to introduce more integration variables than the number of elements in the Cartan of the gauge group $G$ and extremize $F$ under a set of constraints that bring us back down to the number of elements in the Cartan of $G$.  We denote these constrained integration variables by $\mu_i^{(a)}$ where $1 \leq i \leq 2N$ and $a$ ranges from $1$ to the number of gauge group factors.  Explicitly, for the $G = USp(2N) \times SU(2N)^k$ theory we have
\es{FreeOdd}{
  F_A(\mu_i^{(a)}) &= \sum_{i \neq j} \left[  \frac 12  F_V(\mu_i^{(1)} - \mu_j^{(1)} ) + \sum_{a = 2}^{k+1}  F_V(\mu_i^{(a)} - \mu_j^{(a)})  + \frac 12 F_H (\mu_i^{(k+1)} + \mu_j^{(k+1)} ) \right] \\
   &+ \sum_{i, \ j} \left[ \sum_{a=1}^{k} F_H(\mu_i^{(a)} - \mu_j^{(a + 1)} ) \right] + \sum_i \left[\frac 12 F_V(2 \mu_i^{(1)}) + \sum_{a=1}^{k+1} N_f^{(a)} F_H(\mu_i^{(a)} )\right] \,,
 }
under the constraints $\mu_{N+i}^{(1)} = - \mu_i^{(1)}$ as appropriate for an $USp(2N)$ group, and $\sum_i \mu_i^{(a)} = 0$ for all $a \geq 2$, as appropriate for $SU(2N)$.  For the $G = USp(2N) \times SU(2N)^{k-1} \times USp(2N)$ theory we have
\es{FreeEven1}{
  F_B(\mu_i^{(a)}) &= \sum_{i \neq j} \left[  \frac 12  F_V(\mu_i^{(1)} - \mu_j^{(1)} ) + \sum_{a = 2}^{k}  F_V(\mu_i^{(a)} - \mu_j^{(a)}) +  \frac 12  F_V(\mu_i^{(k+1)} - \mu_j^{(k+1)} )   \right] \\
   &+ \sum_{i, \ j} \left[ \sum_{a=1}^{k} F_H(\mu_i^{(a)} - \mu_j^{(a + 1)} ) \right] \\
   &+ \sum_i \left[\frac 12 F_V(2 \mu_i^{(1)}) + \frac 12 F_V(2 \mu_i^{(k+1)}) + \sum_{a=1}^{k+1} N_f^{(a)} F_H(\mu_i^{(a)} )\right] \,,
 }
under the constraints $\mu_{N+i}^{(1)} = - \mu_i^{(1)}$, $\mu_{N+i}^{(k+1)} = - \mu_i^{(k+1)}$ and $\sum_i \mu_i^{(a)} = 0$ for all $2 \leq a \leq k$, as appropriate for $SU(2N)$.  Lastly, for the $SU(2N)^k$ theory:
\es{FreeEven2}{
  F_C(\mu_i^{(a)}) &= \sum_{i \neq j} \left[ \sum_{a = 1}^{k}  F_V(\mu_i^{(a)} - \mu_j^{(a)}) + \frac 12 F_H (\mu_i^{(1)} + \mu_j^{(1)} ) + \frac 12 F_H (\mu_i^{(k)} + \mu_j^{(k)} ) \right] \\
   &+ \sum_{i,\ j} \left[ \sum_{a=1}^{k} F_H(\mu_i^{(a)} - \mu_j^{(a + 1)} )  \right] + \sum_i \left[ \sum_{a=1}^{k} N_f^{(a)} F_H(\mu_i^{(a)} )\right]  \,,
 }
with the constraint that $\sum_i \mu_i^{(a)} = 0$ for all $a$.  We can actually drop the constraints on the $\mu_i^{(a)}$ provided that the extrema of the unconstrained $F(\mu_i^{(a)})$ obey these constraints.  We will see that this is the case, so from now on we forget about these constraints.

Let's assume for simplicity that the $\mu_i^{(a)}$ grow as $N^{1/2}$ at large $N$ as in the previous section.  Finding the extremum of $F(\mu_i^{(a)})$ in the large $N$ approximation is best described as analogous to the first-order degenerate perturbation theory encountered in Quantum Mechanics.  To leading order in $N$, namely $N^{7/2}$ provided that $\mu_i^{(a)} = O(N^{1/2})$ at large $N$, the free energy $F(\mu_i^{(a)})$ is
 \es{FLeading}{
  F_A(\mu_i^{(a)}) &= \frac{\pi}{6} \sum_{i \neq j} \left[  \frac 12  \abs{\mu_i^{(1)} - \mu_j^{(1)}}^3 + \sum_{a = 2}^{k+1}  \abs{\mu_i^{(a)} - \mu_j^{(a)}}^3 - \sum_{a=1}^{k} \abs{\mu_i^{(a)} - \mu_j^{(a + 1)} }^3 - \frac 12 \abs{\mu_i^{(k+1)} + \mu_j^{(k+1)} }^3 \right] \,, \\
    F_B(\mu_i^{(a)}) &= \frac{\pi}{6} \sum_{i \neq j} \left[  \frac 12  \abs{\mu_i^{(1)} - \mu_j^{(1)}}^3 + \sum_{a = 2}^{k}  \abs{\mu_i^{(a)} - \mu_j^{(a)}}^3 + \frac 12  \abs{\mu_i^{(k+1)} - \mu_j^{(k+1)}}^3 - \sum_{a=1}^{k} \abs{\mu_i^{(a)} - \mu_j^{(a + 1)} }^3  \right] \,, \\
    F_C(\mu_i^{(a)}) &= \frac{\pi}{6} \sum_{i \neq j} \left[ \sum_{a = 1}^{k}  \abs{\mu_i^{(a)} - \mu_j^{(a)}}^3 - \sum_{a=1}^{k} \abs{\mu_i^{(a)} - \mu_j^{(a + 1)} }^3 - \frac 12 \abs{\mu_i^{(1)} + \mu_j^{(1)} }^3 - \frac 12 \abs{\mu_i^{(k)} + \mu_j^{(k)} }^3 \right] \,,
 }
where we used \eqref{FVH}.  It is straightforward to see that this expression is extremized provided that all sets of $\mu_i^{(a)}$ are equal and that the $\mu_i^{(a)}$ are distributed symmetrically around $0$.  In other words,
 \es{AllEqual}{
  \mu_i^{(a)} &= \mu_i \,, \qquad  1 \leq i \leq 2 N \,,\\
   \mu_i = -\mu_{N+i} &= \lambda_i \,, \qquad 1 \leq i \leq N \,,
 }
for some $\lambda_i$ which are undetermined yet.  The value of \eqref{FLeading} on the configuration \eqref{AllEqual} is zero.  In other words, the leading order free energy attains its extremum, which so happens to be equal to zero, on any of the degenerate configurations \eqref{AllEqual}.  To find the first correction to \eqref{FLeading} all we have to do is plug in \eqref{AllEqual} into \eqref{FreeOdd}--\eqref{FreeEven2} and minimize each of these functions with respect to the $\lambda_i$.  We do not need to compute the $1/N$ correction to \eqref{AllEqual} in order to find the first correction to the free energy \eqref{FLeading}.

Plugging \eqref{AllEqual} into \eqref{FreeOdd}--\eqref{FreeEven2} we obtain in each of the three cases an expression that in the continuum limit reduces to \eqref{FreeZn} with $N_f = \sum_a N_f^{(a)}$.  The continuum limit is defined as described around~\eqref{DensityDef}.

\section{Gravity dual and entanglement entropy}
\label{ENTANGLEMENT}

The $S^5$ free energy computed using field theoretic methods in the previous section can be matched to a gravity computation using the gravity dual proposed in \cite{Brandhuber:1999np, Bergman:2012kr}.  This background is an extremum of the type IIA supergravity action with non-vanishing Romans mass.  The metric takes the form of a warped product between $AdS_6$ of radius $L$ and half of an $S^4 / \Z_n$ of radius $2 L /3$, where $\Z_n$ acts freely on $S^4$.  In string frame,
 \es{GotMetric}{
  ds^2 = \frac{1}{(\sin \alpha)^{1/3}}
   \left[L^2 \frac{-dt^2 + d\vec{x}^2 + dz^2}{z^2} + \frac {4 L^2}9 \left( d\alpha^2 + \cos^2 \alpha\, ds_{S^3 / \Z_n}^2 \right) \right] \,,
 }
where $d\vec{x}^2 = \sum_{i = 1}^4 (dx^i)^2$ and $ds_{S^3/\Z_n}^2$ is the line element
 \es{dsOrbifold}{
  ds_{S^3 / \Z_n}^2 = \frac 14 \left[d \theta^2 + \sin^2 \theta\, d\phi^2 + \left( d\psi - \cos \theta\, d\phi \right)^2 \right] \,.
 }
The ranges of the four angles are $\alpha \in (0, \pi/2]$, $\theta \in [0, \pi)$, $\phi \in [0, 2 \pi)$, and $\psi \in [0, 4 \pi/ n)$.  When $n=1$, the range of $\alpha$ is only half the range needed to describe a full $S^4$.  The quantization of the four-form flux relates the radius of AdS to the parameters $N$ and $N_f$ of the field theory \cite{Bergman:2012kr}:
 \es{GotL}{
  \frac{L^4}{\ell_s^4} = \frac{18 \pi^2 n N}{8 - N_f} \,.
 }
Of the other supergravity fields, let us write down the dilaton because it will be needed later on:
  \es{Dilaton}{
   e^{-2 \phi} = \frac{3 (8 - N_f)^{3/2} \sqrt{n N} }{2 \sqrt{2} \pi} (\sin \alpha)^{5/3} \,.
  }
As can be seen from \eqref{GotMetric} the whole 10d space is singular at $\alpha = 0$, but it can be argued that away from the singularity the supergravity solution can indeed be trusted at large $N$ \cite{Brandhuber:1999np, Bergman:2012kr}.

 One way of computing the $S^5$ free energy from the gravity dual would be to evaluate the on-shell action of the supergravity solutions.  However, because of the singularity mentioned above, we find that the on-shell Lagrangian appears to be non-integrable.  Perhaps there are contributions from the singularity that resolve this divergence.

 Another way of computing the $S^5$ free energy from the gravity dual is to calculate the entanglement entropy across a three-sphere of radius $R$ and extract the universal part of this entanglement entropy, which was argued to equal minus the free energy on $S^5$ \cite{Casini:2011kv}.  The prescription proposed in \cite{Ryu:2006bv} states that the entanglement entropy across a given surface $\Sigma$ in the boundary theory is proportional to the area in Planck units of a minimal surface whose boundary is fixed to be $\Sigma$.  This prescription was generalized in \cite{Klebanov:2007ws} to gravity duals with non-trivial dilaton profile.  In our case, we should consider an 8d spacelike surface that approaches $S^3$ times the half-$S^4/\Z_n$ at the boundary of $AdS_6$.  The entanglement entropy is \cite{Klebanov:2007ws}
  \es{EntEntropy}{
   S = \frac{2}{(2 \pi)^6 \ell_s^8} \int d^8 x\, e^{- 2 \phi} \sqrt{g } \,,
  }
 where $g$ is the determinant of the induced metric on the 8d surface computed from the 10d string frame metric
  \eqref{GotMetric}.  Going to polar coordinates by writing $d\vec{x}^2 = d \rho^2 + \rho^2 d\Omega_3^2$, we can
   parameterize this surface by $\rho = \rho(z)$ and we should require $\rho(0) = R$.  The surface then wraps all the angles and is contained in a constant time slice.  The integral~\eqref{EntEntropy} becomes
    \es{EntEntropyAgain}{
     S = \frac{3 (nN)^{5/2}}{\pi^3 \sqrt{2(8 - N_f)}} \int \frac{\rho(z)^3 \sqrt{1 + \rho'(z)^2}}{z^4} (\sin \alpha)^{\frac 13} (\cos \alpha)^3 d z \wedge d\alpha \wedge \vol_{S^3} \wedge \vol_{S^3 / \Z_n} \,.
    }
Note that the $\alpha$ integral converges at $\alpha = 0$ despite the singularity of the metric \eqref{GotMetric}.  Performing the angular integrals and using $\Vol(S^3 / \Z_n) = 2 \pi^2 / n$, one obtains
 \es{EntExplicit}{
  S = \frac{27 \pi n^{3/2} N^{5/2}}{5 \sqrt{2(8 - N_f)}} \int dz\, \frac{\rho(z)^3 \sqrt{1 + \rho'(z)^2}}{z^4} \,.
 }
The function $\rho(z)$ that extremizes \eqref{EntExplicit} under the boundary condition $\rho(0) = R$ is $\rho(z) = \sqrt{R^2 - z^2}$.  We compute the area of the minimal surface by integrating from $z = z_\text{min}$ to $z = R$:
 \es{UVDivs}{
  S = \frac{9 \sqrt{2} \pi n^{3/2} N^{5/2}}{5 \sqrt{8 - N_f}} \left[\frac{R^3}{2 z_\text{min}^3}
    - \frac{3 R}{2 z_\text{min}} + 1 \right] \,.
 }
The first two terms in the parenthesis correspond to non-universal UV divergences that should be subtracted away.  The remaining finite part is universal and equals minus the free energy on $S^5$ \cite{Casini:2011kv}, so we conclude that the $S^5$ free energy is
 \es{FreeFromEnt}{
  F = - \frac{9 \sqrt{2} \pi n^{3/2} N^{5/2}}{5 \sqrt{8 - N_f}} \,,
 }
in perfect agreement with the expression \eqref{FOrbi} obtained from the matrix model.

\section*{Acknowledgments}

We thank O.~Bergman for useful discussions.  The work of DLJ was supported in part by the Fundamental Laws
Initiative Fund at Harvard University, the National Science
Foundation Grant No.~1066293, and the hospitality of the Aspen
Center for Physics.  The work of SSP was supported by a Pappalardo
Fellowship in Physics at MIT and by the U.S. Department of Energy
under cooperative research agreement Contract Number
DE-FG02-05ER41360\@.  SSP thanks the Berkeley Center for
Theoretical Physics and the Stanford Institute for Theoretical
Physics for hospitality while this work was in progress.

\bibliographystyle{ssg}
\bibliography{5d}

\end{document}